\documentstyle[12pt]{article}
\makeatletter
\@addtoreset{equation}{subsection}
\makeatother

\begin{document}
\begin{center}
{\Large \bf
Quantum Fields in Curved Spacetime:} \\[0.5cm]
{\large\bf Quantum-Gravitational Nonlocality and
Conservation of Particle Numbers}\\[1.5cm]
{\bf Vladimir S.~MASHKEVICH}\footnote {E-mail:
mash@gluk.apc.org}  \\[1.4cm]
{\it Institute of Physics, National academy
of sciences of Ukraine \\
252028 Kiev, Ukraine} \\[1.4cm]
\vskip 1cm

{\large \bf Abstract}
\end{center}

We argue that the conventional quantum field theory in curved
spacetime has a grave drawback: The canonical commutation
relations for quantum fields and conjugate momenta do not hold.
Thus the conventional theory should be denounced and the related
results revised. A Hamiltonian version of the canonical
formalism for a free scalar quantum field is advanced, and the
fundamentals of an appropriate theory are constructed. The
principal characteristic feature of the theory is
quantum-gravitational nonlocality: The Schr\"odinger field
operator at time t depends on the metric at t in the whole
3-space. It is easily comprehended that the canonical commutation
relations may be fulfilled only if that nonlocality takes place.
Applications to cosmology and black holes are given, the results
being in complete agreement with those of general relativity for
particles in curved spacetime. A model of the universe is advanced,
which is an extension of the Friedmann universe; it lifts the
problem of missing dark matter. A fundamental and shocking result
is the following: There is no particle creation in the case of a
free quantum field in curved spacetime; in particular, neither
the expanding universe nor black holes create particles.

\newpage

\section*{Introduction}

The conventional quantum field theory in curved spacetime [1,2]
is based on the following representation of a scalar quantum
field: $\phi(p)=\sum_{j}\{ f_{j}(p)a_{j}+f^{*}_{j}(p)a_{j}
^{\dag}\}$. Here $p$ is a point of spacetime manifold;
$\{ f_{j}\}$ and $\{ f_{j}^{*}\}$ are
complete sets of positive and negative norm solutions to the
Klein-Gordon, or generalized wave equation $(\Box+m^{2})\chi
=0$; $a_{j}$ and $a_{j}^{\dag}$ are annihilation and creation
operators, $[a_{j},a_{j'}]=0,\;[a_{j},a_{j'}^{\dag}]=\delta
_{jj'}$. In the comoving reference frame, where $p=(t,s)$, the
conjugate momentum is $\pi(p)=\dot \phi(p)=\sum_{j}
\{ \dot f_{j}(p)a_{j}+\dot f_{j}^{*}(p)a_{j}^{\dag} \}$.

The canonical commutation relations are: $[\phi(s,t),\phi(s',t)]
=0,\;[\pi(s,t),\pi(s',t)]=0$,\linebreak $[\phi(s,t),\pi(s',t)]={\rm i}
\delta(s,s')$. We obtain for the commutators: $[\phi(s,t),
\phi(s',t)]=\linebreak\sum_{j}\{ f_{j}(s,t)f_{j}^{*}(s',t)-
f_{j}(s',t)f_{j}^{*}(s,t)\},\;[\pi(s,t),\pi(s',t)]=
\sum_{j}\{ \dot f_{j}(s,t)\dot f_{j}^{*}(s',t)-
\dot f_{j}(s',t)\dot f_{j}^{*}(s,t)\}$,\\$[\phi(s,t),
\pi(s',t)]=\sum_{j}\{ f_{j}(s,t)\dot f_{j}^{*}(s',t)-
\dot f_{j}(s',t)f_{j}^{*}(s,t)\}$. In the generic case of
a time-dependent metric, the canonical commutation relations do not
hold. The reason is that the wave equation $(\Box+m^{2})\phi=0$
is local with respect to the metric: For a given operator
$\phi(s,t)$, it is possible to obtain an arbitrary operator
$\phi(s',t)$ by choosing an appropriate metric, which results
in the violation of the relation $[\phi(s,t),\phi(s',t)]=0$.
The violation leads to disastrous effects: It becomes possible
to introduce an absolute notion of simultaneity.

Thus the conventional theory should be denounced and the related
results revised.

In this paper, a consistent theory for a free scalar quantum field
in curved spacetime is advanced. The basic outline of the theory
is as follows.

Spacetime manifold is $M=T\times S$ where $T$ stands for time and
$S$ for 3-space. In the comoving reference frame, metric $g$ is
of the form $g(t,s)=dt\otimes dt-h_{t},\;t\in T,\;s\in S,\;
h_{t}=h(t,s)$.

First and foremost, the canonical commutation relations must be
fulfilled, so that we put in any picture $\phi(t,s)=\frac{1}
{\sqrt{2}}\sum_{j}\frac{1}{\sqrt{\omega_{j}(t)}}\{
u_{j}(t,s)a_{jt}+u_{j}^{*}(t,s)a_{jt}^{\dag} \},\;\pi(t,s)
=\frac{{\rm i}}{\sqrt{2}}\sum_{j}\sqrt{\omega_{j}(t)}\linebreak
\times\{
-u_{j}(t,s)a_{jt}+u_{j}^{*}(t,s)a_{jt}^{\dag}\}$ where
$\{ u_{j}\}$ is a complete set of functions on $S$,
such that $(u_{j},u_{j'})=(u_{j},u_{j'})_{t}
\equiv \int_{S}ds\sqrt{(h_{t})}u_{j}^{*}
u_{j'}=\delta_{jj'},\;(h)={\rm det}(h_{ik}),\;u_{j}^{*}=u_{p(j)},\;p$ is
a permutation, and $\omega_{p(j)}=\omega_{j}$. Now the canonical
commutation relations do hold.

We choose the functions $u_{j}$ to be solutions to the equation
$\Delta u_{j}=-k_{j}^{2}u_{j},\;\Delta \chi=\Delta_{t}(s)\chi=
\frac{1}{\sqrt{(h)}}\partial_{i}[\sqrt{(h)}h^{ik}\partial_{k}
\chi]$, and put $\omega_{j}(t)=(m^{2}+k_{jt}^{2})^{1/2}$. Then
the Hamiltonian is $H_{t}=\sum_{j}\omega_{j}(t)a_{j}^{\dag}a_{j}$.

The principal characteristic feature of the theory is
quantum-gravitational nonlocality: The Schr\"odinger field
operator $\phi_{S}(t,s)$ at time $t$ depends on metric $h_{t}(s)$
in the whole 3-space $S$. In view of the failure of the
conventional theory considered above, it is easily comprehended
that the canonical commutation relations may be fulfilled only
if the nonlocality is involved.

Applications to cosmology and black holes are given. Bearing the
relation $\omega_{j}=(m^{2}+k_{j}^{2})^{1/2}$ in mind, it is
apparent that the results are in complete agreement with those
of general relativity for particles in curved spacetime. A model
of the universe is advanced, which is an extension of the
Friedmann universe; the model lifts the problem of missing dark
matter.

The expression $H_{t}=\sum_{j}\omega_{j}(t)N_{j},\; N_{j}=
a_{j}^{\dag}a_{j}$, for the Hamiltonian in the comoving
reference frame implies the fundamental and shocking result:
There is no particle creation in the case of a free quantum
field in curved spacetime; in particular, neither the expanding
universe nor black holes create particles.

\section{Preliminaries}

\subsection{Spacetime}

The employment of the comoving reference frame implies that
spacetime manifold $M$ is a trivial bundle [3], so that we assume
from the outset that
\begin{equation}
M=T\times S,\quad M\ni p=(t.s),\quad t\in T,\quad s\in S,
\label{1.1.1}
\end{equation}
holds, where $T$ is time and $S$ is 3-space.

Metric $g$ in the comoving reference frame is of the form
\begin{equation}
g=g(t,s)=dt\otimes dt-h_{t}=(dt)^{2}-h_{ik}(t,x)dx^{i}dx^{k}.
\label{1.1.2}
\end{equation}

\subsection{Classical field dynamics}

The Lagrangian density for a real free scalar field $\varphi$
is
\begin{equation}
{\cal L}=\frac{1}{2}\{\partial_{\mu}\varphi\partial^{\mu}
\varphi
-m^{2}\varphi^{2}\}=\frac{1}{2}\{\partial_{t}\varphi
\partial_{t}\varphi-h^{ik}\partial_{i}\varphi\partial_{k}
\varphi-m^{2}\varphi^{2}\}.
\label{1.2.1}
\end{equation}
The related dynamical equation is the Klein-Gordon, or
generalized wave equation,
\begin{equation}
(\Box+m^{2})\varphi=0,
\label{1.2.2}
\end{equation}
\begin{equation}
\Box=\nabla_{\mu}\nabla^{\mu},\quad
\Box\chi=\frac{1}{\sqrt{(h)}}\partial_{t}[\sqrt{(h)}
\partial_{t}\chi]-\Delta\chi,\quad \Delta\chi=
\frac{1}{\sqrt{(h)}}\partial_{i}[\sqrt{(h)}h^{ik}
\partial_{k}\chi],\quad (h)={\rm det}(h_{ik}).
\label{1.2.3}
\end{equation}
The conjugate momentum is\begin{equation}
\pi=\dot\varphi\equiv\partial_{t}\varphi.
\label{1.2.4}
\end{equation}
The Hamiltonian is
\begin{equation}
H_{t}=\frac{1}{2}\int_{S}ds\sqrt{(h_{t})}\{\pi^{2}+
h^{ik}\partial_{i}\varphi\partial_{k}\varphi+
m^{2}\varphi^{2}\}.
\label{1.2.5}
\end{equation}

\subsection{Canonical commutation relations}

The canonical commutation relations for the quantum field
and conjugated momentum operators $\phi,\;\pi$ are of the form
\begin{equation}
[\phi(s,t),\phi(s',t)]=0,
\label{1.3.1}
\end{equation}
\begin{equation}
[\pi(s,t),\pi(s',t)]=0,
\label{1.3.2}
\end{equation}
\begin{equation}
[\phi(s,t),\pi(s',t)]={\rm i}\delta_{t}(s,s')
\label{1.3.3}
\end{equation}
where the delta function $\delta_{t}(s,s')$ is defined by
\begin{equation}
\int_{S}ds\sqrt{(h_{t})}\delta_{t}(s,s')\chi(s)=\chi(s'),\quad
\int_{S}ds'\sqrt{(h_{t})}\delta_{t}(s,s')\chi(s')=\chi(s).
\label{1.3.4}
\end{equation}

\section{The conventional theory and its inconsistency}

\subsection{The standard quantization}

In the conventional theory, the (symplectic [4]) inner product
of solutions to the wave equation (\ref{1.2.2}),
\begin{equation}
(\varphi_{1},\varphi_{2})_{\Omega}=
{\rm i}\int_{S}ds\sqrt{(h)}\left\{
\varphi_{2}^{*}(s,t)\frac{\partial \varphi_{1}(s,t)}
{\partial t }-\varphi_{1}(s,t)\frac{\partial
\varphi_{2}^{*}(s,t)}{\partial t} \right\},
\label{2.1.1}
\end{equation}
is introduced, the related norm being
$(\varphi,\varphi)_{\Omega}$.

Let $\{f_{j}\}$ be a complete set of positive norm solutions
to the wave equation (\ref{1.2.2}); then $\{f_{j}^{*}\}$ will
be a complete set of negative norm solutions, and $\{f_{j},
f_{j}^{*}\}$ form a complete set of solutions. A scalar
quantum field $\phi$ is represented as follows:
\begin{equation}
\phi(s,t)=\sum_{j}\{f_{j}(s,t)a_{j}+f_{j}^{*}(s,t)a_{j}^{\dag}\},
\label{2.1.2}
\end{equation}
where
\begin{equation}
[a_{j},a_{j'}]=0,\quad [a_{j}^{\dag},a_{j'}^{\dag}]=0,\quad
[a_{j},a_{j'}^{\dag}]=\delta_{jj'},
\label{2.1.3}
\end{equation}
and\begin{equation}
\frac{da_{j}}{dt}=0.
\label{2.1.4}
\end{equation}
In view of eqs.(\ref{1.2.4}),(\ref{2.1.4}), the conjugate momentum is
\begin{equation}
\pi(s,t)=\dot \phi(s,t)=\sum_{j}\{\dot f_{j}(s,t)a_{j}+
\dot f_{j}^{*}(s,t)a_{j}^{\dag}\}.
\label{2.1.5}
\end{equation}
The commutators for $\phi$ and $\pi$ take the following form:
\begin{equation}
[\phi(s,t),\phi(s',t)]=\sum_{j}\{f_{j}(s,t)f_{j}^{*}(s',t)
-f_{j}(s',t)f_{j}^{*}(s,t)\},
\label{2.1.6}
\end{equation}
\begin{equation}
[\pi(s,t),\pi(s',t)]=\sum_{j}\{\dot f_{j}(s,t)\dot f_{j}^{*}
(s',t)-\dot f_{j}(s',t)\dot f_{j}^{*}(s,t)\},
\label{2.1.7}
\end{equation}
\begin{equation}
[\phi(s,t),\pi(s',t)]=\sum_{j}\{f_{j}(s,t)\dot f_{j}^{*}(s',t)
-\dot f_{j}(s',t)f_{j}^{*}(s,t)\}.
\label{2.1.8}
\end{equation}

\subsection{The Wald quantization}

Let us consider the quantization advanced by Wald [4]. For
every solution $\varphi$ to the wave equation (\ref{1.2.2}), by
eq.([4].3.2.27), the relation
\begin{equation}
\varphi(f)=\Omega(Ef,\varphi)
\label{2.2.1}
\end{equation}
holds, where $\varphi(f)$ is a smeared field, $Ef$ is a solution
to the wave equation, and $\Omega$ is the symplectic structure
given by eq.([4].4.2.6),which is equivalent to the inner product
(\ref{2.1.1}). The smeared Heisenberg field operator ([4].4.2.9)
is
\begin{equation}
\phi(f)={\rm i}b_{f}^{\dag}\equiv{\rm i}\{ a
(\overline{K(Ef)})-a^{\dag}(K(Ef))\},
\label{2.2.2}
\end{equation}
where $K(Ef)$ is a vector of the Hilbert space. By eq.([4].3.2.30),
the commutation relation
\begin{equation}
[\phi(f),\phi(g)]=-{\rm i}\Omega(Ef,Eg)
\label{2.2.3}
\end{equation}
holds.

We have the following relations:
\begin{equation}
\phi(f^{*})=[\phi(f)]^{\dag}=-{\rm i}b_{f},
\label{2.2.4}
\end{equation}
\begin{equation}
[b_{f},b_{g}]={\rm i}\Omega(Ef^{*},Eg^{*}),\quad [b_{f}^{\dag},
b_{g}^{\dag}]={\rm i}\Omega(Ef,Eg),\quad [b_{f},b_{g}^{\dag}]=
-{\rm i}\Omega(Ef^{*},Eg).
\label{2.2.5}
\end{equation}
Let, for a complete set $\{Ef_{j},Ef_{j}^{*}\}$ of the solutions
to the wave equation, the relations
\begin{equation}
\Omega(Ef_{j},Ef_{j'})=0,\quad \Omega(Ef_{j}^{*},Ef_{j'}^{*})=0,
\quad \Omega(Ef_{j}^{*},Ef_{j'})={\rm i}\delta_{jj'},
\quad \Omega(Ef_{j},Ef_{j'}^{*})=-{\rm i}\delta_{jj'}
\label{2.2.6}
\end{equation}
hold; then we obtain
\begin{equation}
[b_{j},b_{j'}]=0,\quad [b_{j}^{\dag},b_{j'}^{\dag}]=0,
\quad [b_{j},b_{j'}^{\dag}]=\delta_{jj'}.
\label{2.2.7}
\end{equation}
We put
\begin{equation}
\phi=\sum_{j}\{(Ef_{j})b_{j}+(Ef_{j}^{*})b_{j}^{\dag}\},
\label{2.2.8}
\end{equation}
then
\begin{equation}
\phi(f_{j})={\rm i}b_{j}^{\dag},\quad \phi(f_{j}^{*})=
-{\rm i}b_{j},
\label{2.2.9}
\end{equation}
which corresponds to eqs.(\ref{2.2.2}),(\ref{2.2.4}). The
representation (\ref{2.2.8}) is equivalent to the standard
representation (\ref{2.1.2}).

\subsection{The problem of commutation relations}

In view of eq.(\ref{2.1.6}), the canonical commutation relation
(\ref{1.3.1}) implies that the equality
\begin{equation}
\sum_{j}f_{j}(s,t)f_{j}^{*}(s',t)=\sum_{j}
f_{j}(s',t)f_{j}^{*}(s,t)
\label{2.3.1}
\end{equation}
must hold. Let us write the equality as
\begin{equation}
\sum_{j}F_{j}(s,s')=\sum_{j}F_{j}(s',s),\qquad F_{j}(s,s')
\ne F_{j}(s',s).
\label{2.3.2}
\end{equation}
Generally, we have
\begin{equation}
F_{j}(s',s)=F_{p(j)}(s,s')
\label{2.3.3}
\end{equation}
where $p$ is a permutation, such that
\begin{equation}
p\circ p={\rm I},\quad p^{-1}=p.
\label{2.3.4}
\end{equation}
Thus we obtain
\begin{equation}
f_{j}(s',t)f_{j}^{*}(s,t)=f_{p(j)}(s,t)f_{p(j)}^{*}(s',t),
\label{2.3.5}
\end{equation}
whence
\begin{equation}
\frac{f_{j}^{*}(s,t)}{f_{p(j)}(s,t)}=
\frac{f_{p(j)}^{*}(s',t)}{f_{j}(s',t)}=z_{j}(t)=z_{p(j)}(t).
\label{2.3.6}
\end{equation}
For $s'=s$ we obtain
\begin{equation}
|f_{p(j)}|^{2}=|f_{j}|^{2},\qquad |z_{j}|=1,\quad z_{j}=
e^{{\rm i}2\alpha_{j}(t)},
\label{2.3.7}
\end{equation}
so that
\begin{equation}
f_{j}(s,t)=e^{-\alpha_{j}(t)}f_{j}^{0}(s,t),\quad
f_{j}^{*}(s,t)=e^{{\rm i}\alpha_{j}(t)}f_{j}^{0*}(s,t),
\qquad f_{j}^{0*}=f_{p(j)}^{0}.
\label{2.3.8}
\end{equation}
In view of eqs.(\ref{2.1.8}),(\ref{1.3.3}),
\begin{equation}
\frac{d\alpha_{j}}{dt}\ne 0.
\label{2.3.9}
\end{equation}
Similarly, from eqs.(\ref{2.1.7}),(\ref{1.3.2}) we obtain
\begin{equation}
e^{-{\rm i}\beta_{j}}\dot f_{j}^{*}=-
e^{{\rm i}\beta_{j}}\dot f_{p(j)},\quad \beta_{j}=\beta_{j}(t),
\label{2.3.10}
\end{equation}
so that
\begin{equation}
e^{-{\rm i}\beta_{j}}\frac{\partial}{\partial t}
\left[ e^{{\rm i}\alpha_{j}}f_{j}^{0*} \right]=
-e^{{\rm i}\beta_{j}}\frac{\partial}{\partial t}
\left[ e^{-{\rm i}\alpha_{j}}f_{j}^{0*} \right],
\label{2.3.11}
\end{equation}
whence generally
\begin{equation}
\beta_{j}=\alpha_{j},\qquad \frac{\partial f_{j}^{0*}}
{\partial t}=0,
\label{2.3.12}
\end{equation}
and
\begin{equation}
f_{j}(s,t)=e^{-{\rm i}\alpha_{j}(t)}f_{j}^{0}(s).
\label{2.3.13}
\end{equation}
But if
\begin{equation}
\frac{\partial h}{\partial t}\ne 0,
\label{2.3.14}
\end{equation}
solutions to the wave equation (\ref{1.2.2}) are not of the form
\begin{equation}
\varphi(s,t)=u(t)v(s).
\label{2.3.15}
\end{equation}
Thus in the generic case of a nonstationary metric, the canonical
commutation relations do not hold. The reason is that the wave
equation is local with respect to the metric: For a given operator
$\phi(s,t)$, it is possible to obtain an arbitrary operator
$\phi(s',t)$ by choosing an appropriate metric $h_{t}$, which
results in the violation of the relation (\ref{1.3.1}).

Note that the commutation relations (\ref{2.1.3}) are worthwhile
if and only if they imply the canonical commutation relations
(\ref{1.3.1})-(\ref{1.3.3}), which is not the case in the
conventional theory.

The violation of eq.(\ref{1.3.1}) leads to disastrous effects.
In view of the uncertainty relation
\begin{equation}
\Delta_{\Psi}\phi(s,t)\Delta_{\Psi}\phi(s',t)\ge \frac{1}{2}
|(\Psi,[\phi(s,t),\phi(s',t)]\Psi)|,
\label{2.3.16}
\end{equation}
measuring $\phi(s',t)$ results in
\begin{equation}
\Delta\phi(s,t)=\infty,
\label{2.3.17}
\end{equation}
i.e., in a prodigious quantum nonlocality. This nonlocality
makes it possible to synchronize clocks at points $s$ and $s'$
in an absolute way or, what is the same, to introduce an absolute
notion of simultaneity.

Let $\phi(s,t)$ be measured quasicontinuously,
\begin{equation}
{\rm for}\quad  t\le t_{0}\qquad \Psi=\Psi_{s},\quad \phi(s,t)\Psi_{s}=
\xi_{s}(t)\Psi_{s},
\label{2.3.18}
\end{equation}
and $\phi(s',t)$ be measured at $t_{0}$. Then the value of $\xi_{s}$
changes by a jump at $t_{0}$. Note that the effect is absent for
$\partial h/\partial t=0$ but does not vanish in the limit
$\partial h/\partial t\to 0$.

We conclude that the conventional theory should be denounced.

\section{Hamiltonian version of the canonical formalism}

We shall be based on the Hamiltonian version of the canonical
formalism, which is the most reliable [5].

\subsection{The Schr\"odinger picture}

In view of eq.(\ref{1.2.5}), we adopt the Schr\"odinger
Hamiltonian
\begin{equation}
\begin{array}{l}
H_{St}=\frac{1}{2}\int_{S}ds\sqrt{(h_{t})}
\left\{ \pi_{S}^{2}+h^{ik}\partial_{i}\phi_{S}
\partial_{k}\phi_{S}+m^{2}\phi_{S}^{2} \right\}\\
\qquad {}=\frac{1}{2}\int_{S}ds \sqrt{(h_{t})}
\left\{ \pi_{S}^{2}-\phi_{S}\Delta\phi_{S}
+m^{2}\phi_{S}^{2} \right\}
\end{array}
\label{3.1.1}
\end{equation}
where $\phi_{S}$ and $\pi_{S}$ are the Schr\"odinger operators
for the field and conjugate momentum respectively.

The standard scalar product is defined by
\begin{equation}
(\chi_{1},\chi_{2})=(\chi_{1},\chi_{2})_{t}=
\int_{S}ds\sqrt{(h_{t})}\chi_{1}^{*}(s)\chi_{2}(s).
\label{3.1.2}
\end{equation}
Let $\{u_{j}(s,t)\}$ be a complete set on $S$ for every $t\in T$,
such that
\begin{equation}
(u_{j},u_{j'})_{t}=\delta_{jj'},
\label{3.1.3}
\end{equation}
\begin{equation}
u_{j}^{*}=u_{p(j)},\quad u_{j}=u_{p(j)}^{*}=u_{(p\circ p)(j)},
\quad p\circ p={\rm I},\quad p^{-1}=p,\quad \sum_{p(j)}=
\sum_{j}.
\label{3.1.4}
\end{equation}
We put
\begin{equation}
\phi_{S}(s,t)=\frac{1}{\sqrt{2}}\sum_{j}\frac{1}
{\sqrt{\omega_{j}(t)}}\{ u_{j}(s,t)a_{jS}
+u_{j}^{*}(s,t)a_{jS}^{\dag}\},
\label{3.1.5}
\end{equation}
\begin{equation}
\pi_{S}(s,t)=\frac{{\rm i}}{\sqrt{2}}\sum_{j}
\{-u_{j}(s,t)a_{jS}+u_{j}^{*}(s,t)a_{jS}^{\dag}\},
\label{3.1.6}
\end{equation}
with
\begin{equation}
\omega_{p(j)}=\omega_{j},
\label{3.1.7}
\end{equation}
and
\begin{equation}
\frac{da_{jS}}{dt}=0,\qquad [a_{jS},a_{j'S}]=0,\quad
[a_{jS}^{\dag},a_{j'S}^{\dag}]=0,\quad [a_{jS},a_{j'S}^{\dag}]
=\delta_{jj'}.
\label{3.1.8}
\end{equation}
We find in any picture
\begin{equation}
\begin{array}{l}
[\phi(s,t),\phi(s',t)]=[\phi_{S}(s,t),\phi_{S}(s',t)]\\
\qquad {}=\frac{1}{2}\sum_{j}\frac{1}{\omega_{j}}
\{u_{j}(s,t)u_{j}^{*}(s',t)-u_{j}^{*}(s,t)u_{j}(s',t)\}\\
\qquad {}=\frac{1}{2}\sum_{j}\frac{1}{\omega_{j}}
\{u_{j}(s,t)u_{j}^{*}(s,'t)-u_{p(j)}(s,t)u_{p(j)}^{*}(s',t)\}\\
\qquad {}=\frac{1}{2}\sum_{j}\frac{1}{\omega_{j}}u_{j}(s,t)u_{j}^{*}(s',t)
-\frac{1}{2}\sum_{p(j)}\frac{1}{\omega_{p(j)}}
u_{p(j)}(s,t)u_{p(j)}^{*}(s',t)=0,
\end{array}
\label{3.1.9}
\end{equation}
\begin{equation}
[\pi(s,t),\pi(s',t)]=-\frac{1}{2}\sum_{j}\omega_{j}
\{u_{j}(s,t)u_{j}^{*}(s',t)-u_{j}^{*}(s,t)u_{j}(s',t)\}=0,
\label{3.1.10}
\end{equation}
\begin{equation}
\begin{array}{l}
[\phi(s,t),\pi(s',t)]=\frac{{\rm i}}{2}
\sum_{j}\{u_{j}(s,t)u_{j}^{*}(s',t)+u_{j}^{*}(s,t)u_{j}(s',t)\}
={\rm i}\sum_{j}u_{j}(s,t)u_{j}^{*}(s',t)\\
\qquad {}={\rm i}\delta_{t}(s,s').
\end{array}
\label{3.1.11}
\end{equation}

We have seen that the relations (\ref{3.1.8}) imply the relations
(\ref{1.3.1})-(\ref{1.3.3}). It easy to see that the reverse is
true as well. We introduce operators
\begin{equation}
\phi_{j}=(u_{j},\phi),\quad \phi_{j}^{\dag}=(\phi,u_{j}),\quad
\pi_{j}=(u_{j},\pi),\quad \pi_{j}^{\dag}=(\pi,u_{j}).
\label{3.1.12}
\end{equation}
It follows from the relations (\ref{1.3.1})-(\ref{1.3.3}) that
\begin{equation}
\begin{array}{l}
[\phi_{j},\phi_{j'}]=0,\quad [\phi_{j},\phi_{j'}^{\dag}]=0,
\quad [\pi_{j},\pi_{j'}]=0,\quad [\pi_{j},\pi_{j'}^{\dag}]=0,\\
\qquad {}[\phi_{j},\pi_{j'}]={\rm i}(u_{j},u_{j'}^{*})=
{\rm i}\delta_{jp(j')},\quad [\phi_{j},\pi_{j'}^{\dag}]=
{\rm i}(u_{j},u_{j'})={\rm i}\delta_{jj'}.
\end{array}
\label{3.1.13}
\end{equation}
We find from eqs.(\ref{3.1.5}),(\ref{3.1.6})
\begin{equation}
a_{j}=\frac{1}{\sqrt{2}}\left\{ \sqrt{\omega_{j}}
\phi_{j}+\frac{{\rm i}}{\sqrt{\omega_{j}}}\pi_{j} \right\},
\quad a_{j}^{\dag}=\frac{1}{\sqrt{2}}\left\{
\sqrt{\omega_{j}}\phi_{j}^{\dag}-\frac{{\rm i}}
{\sqrt{\omega_{j}}}\pi_{j}^{\dag} \right\}.
\label{3.1.14}
\end{equation}
Now eqs.(\ref{3.1.14}),(\ref{3.1.13}) result in the
commutation relations (\ref{3.1.8}).

Thus the commutation relations (\ref{3.1.8}) and
(\ref{1.3.1})-(\ref{1.3.3}) are equivalent to each other.

We have for the Schr\"odinger field and momentum
\begin{equation}
\begin{array}{l}
[\phi_{S}(s_{1},t_{1}),\phi_{S}(s_{2},t_{2})]=\frac{1}{2}
\sum_{j}\frac{1}{\sqrt{\omega_{j}(t_{1})\omega_{j}(t_{2})}}
\{u_{j}(s_{1},t_{1})u_{j}^{*}(s_{2},t_{2})-
u_{j}^{*}(s_{1},t_{1})u_{j}(s_{2},t_{2})\}\\
\qquad {}=\frac{1}{2}\sum_{j}\frac{1}{\sqrt{\omega_{j}(t_{1})
\omega_{j}(t_{2})}}u_{j}(s_{1},t_{1})u_{j}^{*}(s_{2},t_{2})-
\frac{1}{2}\sum_{j}\frac{1}{\sqrt{\omega_{p(j)}(t_{1})
\omega_{p(j)}(t_{2})}}u_{p(j)}(s_{1},t_{1})
u_{p(j)}^{*}(s_{2},t_{2})=0,
\end{array}
\label{3.1.15}
\end{equation}
\begin{equation}
[\pi_{S}(s_{1},t_{1}),\pi_{S}(s_{2},t_{2})]=0.
\label{3.1.16}
\end{equation}

Let ${u_{j}}$ be determined by
\begin{equation}
\Delta u_{j}=-k_{j}^{2}u_{j},\qquad k_{p(j)}^{2}=
k_{j}^{2}=k_{j}^{2}(t).
\label{3.1.17}
\end{equation}
We obtain
\begin{equation}
\begin{array}{l}
H_{St}=\frac{1}{4}\sum_{j}\{
[-\sqrt{\omega_{j}\omega_{p(j)}}+\frac{1}{
\sqrt{\omega_{j}\omega_{p(j)}}}(m^{2}+k_{j}^{2})]
(a_{p(j)S}a_{jS}+a_{p(j)S}^{\dag}a_{jS}^{\dag})\\
\qquad {}+[\omega_{j}+\frac{1}{\omega_{j}}(m^{2}+k_{j}^{2})]
(a_{jS}a_{jS}^{\dag}+a_{jS}^{\dag}a_{jS})\}.
\end{array}
\label{3.1.18}
\end{equation}
We put
\begin{equation}
\omega_{j}=(m^{2}+k_{j}^{2})^{1/2}=\omega_{j}(t),\qquad
\omega_{p(j)}=\omega_{j},
\label{3.1.19}
\end{equation}
then
\begin{equation}
H_{St}=\frac{1}{2}\sum_{j}\omega_{j}(t)(a_{jS}a_{jS}^{\dag}
+a_{jS}^{\dag}a_{jS}).
\label{3.1.20}
\end{equation}
Normal ordering produces
\begin{equation}
H_{St}=\sum_{j}\omega_{j}(t)a_{jS}^{\dag}a_{jS}.
\label{3.1.21}
\end{equation}

The Schr\"odinger equation
\begin{equation}
\frac{d\Psi_{St}}{dt}=-{\rm i}H_{St}\Psi_{St}
\label{3.1.22}
\end{equation}
yields
\begin{equation}
\Psi_{St}=U(t,t_{0})\Psi_{St_{0}},
\label{3.1.23}
\end{equation}
where, with regard to
\begin{equation}
[H_{St_{1}},H_{St_{2}}]=0,
\label{3.1.24}
\end{equation}
\begin{equation}
U(t,t_{0})=\exp\left\{ -{\rm i}\int_{t_{0}}^{t}H_{St'}
dt'\right\}=\prod_{j}\exp\{-{\rm i}\alpha_{j}(t,t_{0})
a_{jS}^{\dag}a_{jS}\},
\label{3.1.25}
\end{equation}
\begin{equation}
\alpha_{j}(t,t_{0})=\int_{t_{0}}^{t}\omega_{j}(t')dt'.
\label{3.1.26}
\end{equation}

\subsection{The Heisenberg picture}

In the Heisenberg picture, we have
\begin{equation}
\Psi_{H}=U(t_{0},t)\Psi_{St}=\Psi_{St_{0}},
\label{3.2.1}
\end{equation}
\begin{equation}
A_{Ht}=U(t_{0},t)A_{St}U(t,t_{0}),
\label{3.2.2}
\end{equation}
so that
\begin{equation}
a_{jHt}=e^{-{\rm i}\alpha_{j}(t,t_{0})}a_{jS},\quad
a_{jHt}^{\dag}=e^{{\rm i}\alpha_{j}(t,t_{0})}a_{jS}^{\dag},
\label{3.2.3}
\end{equation}
and
\begin{equation}
\begin{array}{l}
\phi_{H}(s,t)=\frac{1}{\sqrt{2}}\sum_{j}\frac{1}
{\sqrt{\omega_{j}(t)}}\{u_{j}(s,t)a_{jHt}
+u_{j}^{*}(s,t)a_{jHt}^{\dag}\}\\
\qquad {}=\frac{1}{\sqrt{2}}\sum_{j}\frac{1}{\sqrt
{\omega_{j}(t)}}\{\tilde u_{j}(s,t)a_{jS}+
\tilde u_{j}^{*}(s,t)a_{jS}^{\dag}\},
\end{array}
\label{3.2.4}
\end{equation}
\begin{equation}
\tilde u_{j}(s,t)=e^{-{\rm i}\alpha_{j}(t,t_{0})}u_{j}(s,t).
\label{3.2.5}
\end{equation}
The Hamiltonian
\begin{equation}
H_{Ht}=H_{St}=H_{t}=\sum_{j}\omega_{j}(t)a_{jS}^{\dag}a_{jS}.
\label{3.2.6}
\end{equation}
The equations of motion are
\begin{equation}
\frac{\partial A_{Ht}}{\partial t}=\left
(\frac{\partial A_{St} }{\partial t }\right)_{H}+
{\rm i}[H_{t},A_{Ht}].
\label{3.2.7}
\end{equation}

We have
\begin{equation}
\begin{array}{l}
[\phi_{H}(s_{1},t_{1}),\phi_{H}(s_{2},t_{2})]=
\frac{1}{2}\sum_{j}\frac{1}{\sqrt{\omega_{j}(t_{1})
\omega_{j}(t_{2})}}\{\tilde u_{j}(s_{1},t_{1})
\tilde u_{j}^{*}(s_{2},t_{2})-
\tilde u_{j}^{*}(s_{1},t_{1})\tilde u_{j}(s_{2},t_{2})\}\\
\qquad {}={\rm i}\sum_{j}\frac{1}{\sqrt{\omega_{j}(t_{1})
\omega_{j}(t_{2})}}u_{j}(s_{1},t_{1})u_{j}^{*}(s_{2},t_{2})
\sin[\alpha_{j}(t_{2},t_{1})].
\end{array}
\label{3.2.8}
\end{equation}

\subsection{The energy-momentum tensor}

Normal ordering on the standard energy-momentum tensor
[1] produces\begin{equation}
T_{\mu\nu}=:\left\{\partial_{\mu}\phi\partial_{\nu}\phi-
\frac{1}{2}g_{\mu\nu}g^{\rho\sigma}
\partial_{\rho}\phi\partial_{\sigma}\phi+
\frac{1}{2}m^{2}g_{\mu\nu}\phi^{2} \right\}:\;,
\label{3.3.1}
\end{equation}
whence in the comoving reference frame we have
\begin{equation}
T_{00}=\frac{1}{2}:\left\{ \pi^{2}+h^{ik}
\partial_{i}\phi\partial_{k}\phi+m^{2}\phi^{2} \right\}:\;,
\label{3.3.2}
\end{equation}
\begin{equation}
H_{t}=\int_{S}ds\sqrt{(h_{t})}T_{00},
\label{3.3.3}
\end{equation}
\begin{equation}
\begin{array}{l}
T_{ik}=:\left\{\partial_{i}\phi\partial_{k}\phi+
\frac{1}{2}h_{ik}[\pi^{2}-h^{lm}\partial_{l}\phi
\partial_{m}\phi-m^{2}\phi^{2}]  \right\}:\\
\qquad {}=:\partial_{i}\phi\partial_{k}\phi:+
h_{ik}[:\pi^{2}:-T_{00}],
\end{array}
\label{3.3.4}
\end{equation}
and
\begin{equation}
(\Psi,T_{ik}\Psi)=(\Psi,:\partial_{i}\phi\partial_{k}\phi:\Psi)
+h_{ik}(\Psi,[:\pi^{2}:-T_{00}]\Psi).
\label{3.3.5}
\end{equation}

\section{Quantum-gravitational nonlocality. Covariance and geometry}

\subsection{The violation of the wave equation}

With eqs.(\ref{3.2.4}),(\ref{3.2.5}) in mind, we find
\begin{equation}
(\Box+m^{2})\left[ \frac{\tilde u_{j}}{\sqrt{\omega_{j}}}\right]
=\frac{1}{\sqrt{(h)}}\frac{\partial \sqrt{(h)}}{\partial t}
\frac{\partial}{\partial t}\left[ \frac{\tilde u_{j}}
{\sqrt{\omega_{j}}} \right]+\frac{\partial}{\partial t}
\left\{ \frac{\partial}{\partial t}\left[ \frac{u_{j}}
{\sqrt{\omega_{j}}} \right]e^{-{\rm i}\alpha_{j}} \right\}-
{\rm i}\frac{\partial}{\partial t}[\sqrt{\omega_{j}}u_{j}]
e^{-{\rm i}\alpha_{j}}.
\label{4.1.1}
\end{equation}
Let for some $(t,s)$
\begin{equation}
\frac{d\omega_{j}}{dt}=0,\quad \frac{\partial u_{j}}
{\partial t}=0
\label{4.1.2}
\end{equation}
hold, then
\begin{equation}
(\Box+m^{2})\left[ \frac{\tilde u_{j}}{\sqrt{\omega_{j}}} \right]
=-{\rm i}\frac{1}{\sqrt{(h)}}
\frac{\partial \sqrt{(h)}}{\partial t}
\sqrt{\omega_{j}}\tilde u_{j}.
\label{4.1.3}
\end{equation}
Thus the wave equation is violated in the generic case of a
nonstationary metric.

Note that in general relativity, the relation (\ref{3.1.19})
for $\omega_{j}$ results from the geodesic equation,
\begin{equation}
\ddot x^{k}+\Gamma_{ij}^{k}\dot x^{i}\dot x^{j}=0,
\label{4.1.4}
\end{equation}
rather than from the wave equation.

The wave equation being rejected, the equations of motion are
those in the Schr\"odinger and Heisenberg pictures respectively.

\subsection{Quantum-gravitational nonlocality and commutation
relations}

A local change in the metric $h$ results in changing the Laplacian
$\Delta$ and, by the same token, solutions to the equation
(\ref{3.1.17}), i.e., $k_{j}^{2},\;u_{j},\;\omega_{j}$, and
$\tilde u_{j}/\sqrt{\omega_{j}}$. We call this phenomenon
quantum-gravitational nonlocality.

Generally, quantum-gravitational nonlocality means that
\begin{equation}
A_{St}(s)=A_{S}[s;h_{t}],\qquad H_{t}=H[h_{t}],
\label{4.2.1}
\end{equation}
i.e., that a Schr\"odinger operator at time $t$ depends on
the metric $h_{t}$ in the whole 3-space $S$.

The canonical commutation relation (\ref{1.3.1}) cannot hold
if the dependence of $\phi$ on the metric is local: In that case,
for a given operator $\phi(s,t)$ the operator $\phi(s',t)$
might be arbitrary. Thus in order that the canonical commutation
relations hold, quantum-gravitational nonlocality should take place.

A strictly local measurement of $\phi$ affects metric infinitesimally
only and cannot be used for synchronizing clocks.

The degree of quantum-gravitational nonlocality may be given
by the quantity
\begin{equation}
\beta=\frac{\partial }{\partial t}\left.\left[ \frac{u}{\sqrt
{\omega}}\right]\right/\left[ \frac{u}{\sqrt{\omega}}
\right]\omega=\left.\frac{\partial u}{\partial t}\right/u\omega
-\left.\frac{1}{2}\frac{d\omega}{dt}\right/\omega^{2}.
\label{4.2.2}
\end{equation}
We have
\begin{equation}
\frac{d\omega}{dt}=\frac{k}{\omega}\frac{dk}{dt},
\label{4.2.3}
\end{equation}
so that
\begin{equation}
\beta=\left.\frac{\partial u}{\partial t}\right/u\omega-
\frac{k}{2\omega^{3}}\frac{dk}{dt}.
\label{4.2.4}
\end{equation}

\subsection{The principle of covariance and the geometric principle}

Nonlocality is incompatible with the local principle of
covariance. More general than the latter is the geometric
principle [6]: Spacetime structure and dynamical equations should
be phrased in geometric form. The principle of covariance is a
local version of the geometric principle.

\section{Applications to cosmology}

\subsection{The energy-momentum tensor consisted with metric}

Let in eq.(\ref{3.3.5})
\begin{equation}
(\Psi,:\partial_{i}\phi\partial_{k}\phi:\Psi)\propto h_{ik}
\label{5.1.1}
\end{equation}
hold, i.e.,
\begin{equation}
(\Psi,:\partial_{i}\phi\partial_{k}\phi:\Psi)=Ch_{ik}h^{lm}
(\Psi,:\partial_{l}\phi\partial_{m}\phi:\Psi).
\label{5.1.2}
\end{equation}
Since
\begin{equation}
h^{ik}h_{ik}=3,
\label{5.1.3}
\end{equation}
we find
\begin{equation}
C=\frac{1}{3}
\label{5.1.4}
\end{equation}
and by eqs.(\ref{3.3.5}),(\ref{3.3.2})
\begin{equation}
(\Psi,T_{ik}\Psi)=\frac{1}{3}h_{ik}
(\Psi,\{ 2:\pi^{2}:-T_{00}-m^{2}:\phi^{2}:\}\Psi).
\label{5.1.5}
\end{equation}

\subsection{A homogeneous state}

Let $\Psi$ be a homogeneous state, so that
\begin{equation}
\begin{array}{l}
(\Psi,\{2:\pi^{2}:-T_{00}-m^{2}:\phi^{2}:\}\Psi)=
\frac{1}{V}\int_{S}ds\sqrt{(h)}(\Psi,\{2:\pi^{2}:-
T_{00}-m^{2}:\phi^{2}:\}\Psi),\\
\qquad {}V=V_{t}=\int_{S}ds\sqrt{(h_{t})}.
\end{array}
\label{5.2.1}
\end{equation}
We have
\begin{equation}
\int_{S}ds\sqrt{(h)}T_{00}=\sum_{j}\omega_{j}N_{j},
\qquad N_{j}=N_{jH}=N_{jS}=a_{jS}^{\dag}a_{jS},
\label{5.2.2}
\end{equation}
\begin{equation}
\int_{S}ds\sqrt{(h)}:\pi^{2}:=\sum_{j}\omega_{j}N_{j}+
\{aa+a^{\dag}a^{\dag}\},
\label{5.2.3}
\end{equation}
\begin{equation}
\int_{S}ds\sqrt{(h)}:\phi^{2}:=\sum_{j}\frac{1}{\omega_{j}}N_{j}
+\{aa+a^{\dag}a^{\dag}\}.
\label{5.2.4}
\end{equation}

Let
\begin{equation}
N_{j}\Psi=n_{j}\Psi\qquad {\rm for\;all}\;j,
\label{5.2.5}
\end{equation}
then
\begin{equation}
(\Psi,T_{ik}\Psi)=h_{ik}\frac{1}{3V}\sum_{j}\frac{\omega_
{j}^{2}-m^{2}}{\omega_{j}}n_{j}.
\label{5.2.6}
\end{equation}
Thus the pressure is
\begin{equation}
p=\frac{1}{3V}\sum_{j}\frac{\omega_{j}^{2}-m^{2}}{\omega_{j}}n_{j}
=\frac{1}{3V}\sum_{j}\frac{k_{j}^{2}}{\omega_{j}}n_{j},
\label{5.2.7}
\end{equation}
whereas the energy density is
\begin{equation}
\rho=\frac{E}{V}=\frac{1}{V}\sum_{j}\omega_{j}n_{j}.
\label{5.2.8}
\end{equation}

\subsection{The Robertson-Walker spacetime}

For the Robertson-Walker spacetime, the metric is of the form
\begin{equation}
h(s,t)=R^{2}(t)\kappa(s),\qquad {\rm or}\quad h_{ik}=
R^{2}(t)\kappa_{ik},
\label{5.3.1}
\end{equation}
so that we have
\begin{equation}
(h)=(\kappa)R^{6},\quad (\kappa)={\rm det}(\kappa_{ik}),
\quad \sqrt{(h)}=R^{3}\sqrt{(\kappa)},\quad h^{ik}=
\frac{\kappa^{ik}}{R^{2}},
\label{5.3.2}
\end{equation}
and
\begin{equation}
\Delta=\frac{1}{R^{2}}\Delta_{\kappa},\qquad
\Delta_{\kappa}\chi=\frac{1}{\sqrt{(\kappa)}}\partial_{i}
\left[ \sqrt{(\kappa)}\kappa^{ik}\partial_{k}\chi \right].
\label{5.3.3}
\end{equation}
The equation (\ref{3.1.17}) results in
\begin{equation}
\frac{1}{R^{2}(t)}\Delta_{\kappa}u_{j}=-k_{j}^{2}u_{j},
\label{5.3.4}
\end{equation}
so that, in view of eq.(\ref{3.1.3}),
\begin{equation}
\Delta_{\kappa}u_{j}=-\gamma_{j}^{2}u_{j},\quad
\gamma_{j}^{2}={\rm const},\quad k_{j}^{2}(t)=
\frac{\gamma_{j}^{2}}{R^{2}(t)},\quad u_{j}(s,t)=
\frac{1}{R^{3/2}(t)}u_{j}^{0}(s),
\label{5.3.5}
\end{equation}
and
\begin{equation}
\omega_{j}=\left[ m^{2}+\frac{\gamma_{j}^{2}}{R^{2}(t)}
 \right]^{1/2},
\label{5.3.6}
\end{equation}
the last relation being a familiar result of cosmology.

In eq,(\ref{4.2.4}) we obtain
\begin{equation}
u=\frac{u^{0}(s)}{R^{3/2}(t)},\quad k=\frac{\gamma}{R(t)},
\label{5.3.7}
\end{equation}
so that
\begin{equation}
|\beta|=\frac{3}{2\omega}\frac{dR/dt}{R}
-\frac{1}{2}\frac{(\gamma/R)^{2}}{\omega^{3}}
\frac{dR/dt}{R}=\frac{3H}{2\omega}-
\frac{1}{2}\frac{k^{2}}{\omega^{3}}H=
\left(3-\frac{k^{2}}{\omega^{2}}\right)\frac{H}{2\omega}
<\frac{3H}{2\omega},
\label{5.3.8}
\end{equation}
where $H$ is the Hubble constant.
\begin{equation}
{\rm For}\;H\approx\frac{1}{3}10^{-17}c^{-1}
\;{\rm and}\; \omega\sim 10^{15}c^{-1},\qquad \beta<10^{-32}.
\label{5.3.9}
\end{equation}

With eqs.(\ref{5.2.7}),(\ref{5.2.8}) in mind, we have
\begin{equation}
k_{j}^{2}=\frac{b_{j}^{2}}{V^{2/3}},\quad
b_{j}^{2}={\rm const},\qquad \omega_{j}=\left(
m^{2}+\frac{b_{j}^{2}}{V^{2/3}}\right)^{1/2},
\label{5.3.10}
\end{equation}
so that we find\begin{equation}
\frac{dE}{dV}=\frac{d(\rho V)}{dV}=\sum_{j}n_{j}
\frac{d\omega_{j}}{dV}=-\frac{1}{3V}\sum_{j}n_{j}
\frac{k_{j}^{2}}{\omega_{j}}=-p,
\label{5.3.11}
\end{equation}
i.e.,
\begin{equation}
dE=-pdV,
\label{5.3.12}
\end{equation}
which is a standard relation.

\subsection{Universe dynamics}

In this and the next subsections, we follow the papers [6,7].

The $S$-projected Einstein equation yields
\begin{equation}
G_{ik}=8\pi\kappa_{g}(\Psi,T_{ik}\Psi)\quad \Rightarrow\quad
2\ddot R R+\dot R^{2}+1=-8\pi\kappa_{g} pR^{2}
\label{5.4.1}
\end{equation}
where $\kappa_{g}$ is the gravitational constant; eq.(\ref{5.3.12})
amounts to
\begin{equation}
\frac{d(\rho R^{3})}{dR}=-3pR^{2}.
\label{5.4.2}
\end{equation}
We obtain from eqs.(\ref{5.4.1}),(\ref{5.4.2})
\begin{equation}
\frac{d}{dR}\left(R\dot R^{2}+R-
\frac{8\pi\kappa_{g}}{3}\rho R^{3}    \right)=0,
\label{5.4.3}
\end{equation}
whence
\begin{equation}
R\dot R^{2}+R-\frac{8\pi\kappa_{g}}{3}\rho R^{3}=L=
{\rm const}.
\label{5.4.4}
\end{equation}

The length $L$, which is an integral of motion, is called
cosmic length. In accordance with this, the model considered
is called the cosmic length universe.

The Friedmann universe corresponds to a particular value of
the cosmic length,
\begin{equation}
L_{{\rm Friedmann}}=0.
\label{5.4.5}
\end{equation}
In this sense, the Friedmann universe is the zero-length universe.

The value $L=0$ results from the equation
\begin{equation}
G_{0\mu}=8\pi\kappa_{g}(\Psi,T_{0\mu}\Psi),
\label{5.4.6}
\end{equation}
which is violated by quantum jumps inherent in the generic
case of interacting quantum fields.

\subsection{Lifting the problem of missing dark matter}

The most important problem facing modern cosmology is that of
the missing dark matter [8]. Most of the mass of galaxies and
an even larger fraction of the mass of clusters of galaxies is
dark. The problem is that even more dark matter is required to
account for the rate of expansion of the universe.

More specifically, for the Friedmann universe, the equation
\begin{equation}
\Omega_{0}=2q_{0}
\label{5.5.1}
\end{equation}
holds, where $\Omega$ is the density parameter,
\begin{equation}
\Omega=\frac{\rho}{\rho_{c}},
\label{5.5.2}
\end{equation}
$\rho_{c}$ is the critical value of $\rho$, $q$ is the
deceleration parameter,
\begin{equation}
q=-\frac{\ddot R R}{\dot R^{2}},
\label{5.5.3}
\end{equation}
and subscript 0 indicates present-day values. In particular,
if $q_{0}>1/2$, the universe is closed and $\rho_{0}>\rho_{c}$.
But observational data give $\Omega_{0}<2q_{0}$. Eq.(\ref{5.5.1})
reduces to
\begin{equation}
\Omega_{0}=1+\frac{1}{R_{0}^{2}H_{0}^{2}}.
\label{5.5.4}
\end{equation}

From eq.(\ref{5.4.4}) we obtain
\begin{equation}
\Omega_{0}=1+\frac{1-L/R_{0}}{R_{0}^{2}H_{0}^{2}}
\label{5.5.5}
\end{equation}
in place of eq.(\ref{5.5.4}). For
\begin{equation}
p_{0}\ll\frac{1}{3}\rho_{0},
\label{5.5.6}
\end{equation}
which is fulfilled, eq.(\ref{5.5.5}) reduces to
\begin{equation}
\Omega_{0}=2q_{0}-\frac{L/R_{0}}{R_{0}^{2}H_{0}^{2}}
\label{5.5.7}
\end{equation}
in place of eq.(\ref{5.5.1}).

Eq.(\ref{5.5.7}) lifts the problem.

\section{An application to black holes}

\subsection{The Lema\^\i tre metric}

In the case of a black hole, the metric in the comoving
reference frame is the Lema\^\i tre metric:
\begin{equation}
h=\frac{1}{[(3/2r_{s})(R-t)]^{2/3}}dR^{2}+r_{s}^{2/3}
\left[ \frac{3}{2}(R-t) \right]^{4/3}(d\theta^{2}+
\sin^{2}\theta d\varphi^{2}),
\label{6.1.1}
\end{equation}
where $r_{s}$ is the Schwarzschild radius. The Schwarzschild
coordinate is
\begin{equation}
r=\left(\frac{3}{2} \right)^{2/3}r_{g}^{1/3}(R-t)^{2/3}.
\label{6.1.2}
\end{equation}

\subsection{Quantum field in the comoving reference frame}

With the equation (\ref{3.1.17}) in mind, we find
\begin{equation}
\Delta\chi\equiv\Delta_{\vec R}\chi=\frac{1}{r^{2}}
\partial_{r}[r^{2}\partial_{r}\chi]+\frac{1}{r^{2}\sin\theta}
\partial_{\theta}[\sin\theta\partial_{\theta}\chi]+
\frac{1}{r^{2}\sin^{2}\theta}\partial_{\varphi}^{2}\chi
=\Delta_{\vec r}\chi
\label{6.2.1}
\end{equation}
where
\begin{equation}
\vec R=(R,\theta,\varphi),\qquad \vec r=(r,\theta,\varphi).
\label{6.2.2}
\end{equation}
Thus eq.(\ref{3.1.17}) reduces to
\begin{equation}
\Delta_{\vec r}u_{j}=-k_{j}^{2}u_{j},
\label{6.2.3}
\end{equation}
whence
\begin{equation}
u_{j}=u_{j}(r,\theta,\varphi)
\label{6.2.4}
\end{equation}
with $r$ given by eq.(\ref{6.1.2}), and
\begin{equation}
\frac{dk_{j}^{2}}{dt}=0,\quad \omega_{j}=\left[ m^{2}+
k_{j}^{2} \right]^{1/2},\quad \frac{d\omega_{j}}{dt}=0.
\label{6.2.5}
\end{equation}
So in the comoving reference frame
\begin{equation}
\omega_{j}={\rm const},\qquad H=\sum_{j}\omega_{j}
a_{jS}^{\dag}a_{jS},\quad \frac{dH}{dt}=0.
\label{6.2.6}
\end{equation}

In eq.(\ref{4.2.4}) we obtain
\begin{equation}
\frac{d\omega}{dt}=0,
\label{6.2.7}
\end{equation}
so that
\begin{equation}
\beta=\left.\frac{\partial u}{\partial t}\right/u\omega.
\label{6.2.8}
\end{equation}
We find from eq.(\ref{6.1.2})
\begin{equation}
\frac{\partial u}{\partial t}=\frac{\partial u}{\partial r}
\left(\frac{r_{s}}{r}\right)^{1/2}.
\label{6.2.9}
\end{equation}
By [9], in view of $\sqrt{h}\sim r^{3/2}$,
\begin{equation}
\left|\frac{\partial u}{\partial r}\right|\sim
\left\{\left(k^{2}+\frac{1}{r^{2}}\right)^{1/2}+
\frac{1}{4r}\right\}|u|,
\label{6.2.10}
\end{equation}
so that
\begin{equation}
|\beta|\sim\frac{1}{\omega}\left\{\left(k^{2}+
\frac{1}{r^{2}}\right)^{1/2}+\frac{1}{4r}\right\}
\left(\frac{r_{s}}{r}\right)^{1/2}.
\label{6.2.11}
\end{equation}
In particular,
\begin{equation}
{\rm for}\; r\gg\lambda=\frac{2\pi}{k}\qquad
|\beta|\sim\left[ \frac{\omega^{2}-m^{2}}{\omega^{2}}
\frac{r_{s}}{r} \right]^{1/2}.
\label{6.2.12}
\end{equation}

\section{No particle creation}

\subsection{Conservation of occupation numbers}

We have in the comoving reference frame
\begin{equation}
H_{t}=\sum_{j}\omega_{j}(t)N_{j},\quad
N_{j}=a_{jS}^{\dag}a_{jS},\quad [H_{t},N_{j}]=0,
\label{7.1.1}
\end{equation}
so that occupation numbers are conserved:
\begin{equation}
\frac{dN_{jH}}{dt}=\frac{dN_{jS}}{dt}=0,
\label{7.1.2}
\end{equation}
which implies that there is no particle creation in the
case of a free quantum field.

In particular, neither the expanding universe nor black
holes create particles.

\section*{Acknowledgment}

I would like to thank Stefan V. Mashkevich for helpful
discussions.


\begin{thebibliography}{9}

\bibitem{1} N.D. Birrell, P.C.W. Davies, Quantum Fields
in Curved Space (Cambridge University Press, 1982).
\bibitem{2} L.H. Ford, {\it Quantum Field Theory in Curved Spacetime}
(gr-qc/9707062, 1997).
\bibitem{3} M. Crampin, F.A.E. Pirani, Applicable Differential
Geometry (Cambridge University Press, 1986)
\bibitem{4} Robert M. Wald, Quantum Field Theory in Curved
Spacetime and Black Hole Thermodynamics (The University
of Chicago Press, 1994).
\bibitem{5} Steven Weinberg, The Quantum Theory of Fields,
Vol.1 (Cambridge University Press, 1995).
\bibitem{6} Vladimir S. Mashkevich, gr-qc/9603022.
\bibitem{7} Vladimir S. Mashkevich, gr-qc/9609035.
\bibitem{8} Steven Weinberg, Dreams of a Final Theory
(Vintage, London etc., 1993).
\bibitem{9} Hans A. Bethe, Edwin E. Salpeter, Quantum Mechanics of
One- and Two-Electron Atoms (Springer-Verlag, 1957).
\end{thebibliography}
\end{document}